\documentclass[fleqn,11pt]{wlscirep}
\usepackage[utf8]{inputenc}
\usepackage[T1]{fontenc}
\usepackage{braket}
\usepackage{siunitx}
\usepackage{dsfont}
\usepackage{textcomp}
\usepackage{gensymb}
\DeclareSIUnit\angstrom{\text {Å}}
\usepackage{bm}
\usepackage{mathtools}
\usepackage{caption}
\usepackage[printwatermark]{xwatermark}

\title{Orbital and spin Edelstein effects in KTaO$_3$(110) two-dimensional electron gases}
\author[1,2]{Hugo Witt}
\author[3]{Aravind Raji}
\author[1,4]{Srijani Mallik}
\author[5]{Börge Göbel}
\author[1]{Luis M. Vicente-Arche}
\author[1]{Sara Varotto}
\author[1,6]{Julien Bréhin}
\author[2,7]{Gerbold Ménard}
\author[8,9]{Raphaël Salazar}
\author[8]{Julien Rault}
\author[8]{François Bertran}
\author[8,10]{Patrick Le Fèvre}
\author[1]{Isabella Boventer}
\author[5]{Ingrid Mertig}
\author[1]{Agnès Barthélémy}
\author[3]{Alexandre Gloter}
\author[5,11,12,(*)]{Annika Johansson}
\author[2]{Nicolas Bergeal}
\author[1,($^{+}$)]{Manuel Bibes}

\affil[1]{Laboratoire Albert Fert, CNRS, Thales, Université Paris-Saclay, 91767 Palaiseau, France}
\affil[2]{Laboratoire de Physique et d’Etude des Matériaux, ESPCI Paris, Université PSL, CNRS, 75005, Paris, France}
\affil[3]{Laboratoire de Physique des Solides, CNRS, Université Paris-Saclay, 91405 Orsay, France}
\affil[4]{Saha Institute of Nuclear Physics, 1/AF, Bidhannagar, Kolkata 700 064, India}
\affil[5]{Institute of Physics, Martin-Luther-Universität Halle-Wittenberg, 06099 Halle, Germany}
\affil[6]{Spintec, Université Grenoble Alpes/CNRS/CEA, IRIG / CEA Grenoble, 17, Rue des Martyrs, 38054 Grenoble Cedex, France}
\affil[7]{Laboratoire de Physique de l’École Normale Supérieure, ENS, Université PSL, CNRS, Sorbonne Université, Université Paris Cité, 75005 Paris, France}
\affil[8]{Synchrotron SOLEIL, L’Orme des Merisiers, Saint-Aubin, BP 48, Gif-sur-Yvette Cedex 91192, France}
\affil[9]{New Technologies-Research Centre, University of West Bohemia, Pilsen, Czech Republic}
\affil[10]{Institut de Physique de Rennes, Université de Rennes, CNRS, 35000, Rennes, France}
\affil[11]{Max Planck Institute of Microstructure Physics, Weinberg 2, 06120 Halle, Germany}
\affil[12]{Halle-Berlin-Regensburg Cluster of Excellence CCE, Germany}

\affil[(*)]{annika.johansson@physik.uni-halle.de}
\affil[($^{+}$)]{manuel.bibes@cnrs-thales.fr}

\begin{abstract}
The orbital Edelstein effect converts an electric field into a non-equilibrium orbital polarization, opening new opportunities for orbitronics. Although signatures of the orbital Edelstein effect have been reported, its microscopic mechanisms and quantitative validation remain underexplored. Here, by directly linking the atomic structure of KTaO$_3$(110) two-dimensional electron gases to both their calculated and measured electronic band dispersions, we predict and provide experimental evidence for an orbital Edelstein effect that largely counterbalances its spin counterpart. Scanning transmission electron microscopy and electron energy-loss spectroscopy resolve the interfacial atomic configuration, which is used as input for density-functional calculations. Angle-resolved photoemission spectroscopy then confirms the resulting band structure, which is fitted by a tight-binding model enabling computation of the spin and orbital Edelstein responses. Harmonic magnetotransport indicates that a $\sim$20 \% contribution from the orbital Edelstein response is necessary to describe the magnitude and anisotropy of the effect. Our results establish KTaO$_3$(110) as a model platform for orbitronics and demonstrate a pathway to generate and harness orbital polarization in quantum oxide systems while also offering new insights into pairing mechanisms in their superconducting state.
\end{abstract}

\begin{document}

\flushbottom
\maketitle


This is the pre-peer reviewed version of the following article: H. Witt, A. Raji, S. Mallik, et al. Advanced Materials (2026): e73899, which has been published in final form at https://doi.org/10.1002/adma.73899. This article may be used for non-commercial purposes in accordance with Wiley Terms and Conditions for Use of Self-Archived Versions.


\section*{Introduction}
Oxide interfaces host electronic phases that do not exist in the bulk, driven by symmetry breaking and atomic reconstructions \cite{hwang_emergent_2012}. A prominent example is the two-dimensional electron gas (2DEG) at SrTiO$_3$ (STO) interfaces, which exhibits superconductivity below $T_c\approx \SI{300}{\milli\kelvin}$  \cite{reyren_superconducting_2007} and Rashba spin–orbit coupling \cite{caviglia_tunable_2010} with coefficients 
$\alpha_R \simeq \SIrange{20}{50}{\milli\electronvolt\angstrom}$. This Rashba state enables charge–spin interconversion through the direct \cite{vaz_determining_2020} and inverse Edelstein effects \cite{lesne_highly_2016,vaz_mapping_2019}. The broken inversion symmetry in STO 2DEGs also enables the equivalent of the spin Edelstein effects for the electron’s orbital angular momentum\cite{levitov_magnetoelectric_1985}, coined the orbital Edelstein effect. In a Rashba state, it manifests from particular textures of orbital angular momentum, typically locked transverse to the momentum $\bold{k}$, just like the spin Edelstein effect arises from chiral spin textures due to spin-momentum locking. Remarkably, in STO 2DEGs the orbital Edelstein effect (OEE) was predicted to be larger than its spin equivalent\cite{johansson_theory_2024}. STO 2DEGs have thus emerged as an interesting source and detector of orbital moment, usable for instance to generate more efficient torques on the magnetization of ferromagnets, as has been reported using the orbital Hall effect of
Cu/CuO$_x$\cite{ding_harnessing_2020} or Cu/AlO$_x$ bilayers\cite{krishnia_quantifying_2024}. Indications on an OEE in STO 2DEGs were recently provided by El Hamdi \emph{et al}\cite{el_hamdi_observation_2023}.

The limitations of STO 2DEGs in terms of their low superconducting $T_c$ and moderate $\alpha_R$ may be overcome in a new family of oxide 2DEGs based on KTaO$_3$ (KTO). Just like STO, KTO is a quantum paraelectric that becomes a high mobility metal upon minute doping\cite{wemple_transport_1965}. However, it is not a superconductor in bulk form\cite{thompson_very_1982}. 2DEGs formed on KTO (001) surfaces are not superconducting either --- aside from an early report with a low $T_c \simeq 40$ mK in ionic-liquid-gated crystals\cite{ueno_discovery_2011} --- but show a large Rashba splitting ($\alpha_R \simeq$ $\SI{300}{\milli\electronvolt\angstrom}$), visible in angle resolved photoemission spectroscopy (ARPES)\cite{varotto_direct_2022} and usable for spin-charge interconversion\cite{vicentearche_spincharge_2021}. Surprisingly however, KTO (111) and KTO (110) 2DEGs are superconducting, with $T_c$  up to $\sim 2.2$ K\cite{liu_two-dimensional_2021} and $\sim 1$ K\cite{chen_two-dimensional_2021}, respectively. KTO (111) 2DEGs have
been the most studied: besides their superconducting behaviour\cite{mallik_superfluid_2022,liu_tunable_2023,arnault_anisotropic_2023,chen_orientation-dependent_2024}, attention has been put to their band structure and spin-orbit properties\cite{chen_orientation-dependent_2024,bareille_two-dimensional_2015,bruno_band_2019,mallik_electronic_2023,zhai_large_2023,xu_giant_2024}. The Rashba coefficient is lower than in KTO(001)\cite{mallik_electronic_2023}, but spin-charge conversion has been reported\cite{al-tawhid_spin--charge_2025}.

Here, we focus on KTO(110) 2DEGs generated by depositing an ultrathin Al layer onto a KTO(110) single-crystal substrate. Just as for KTO(001)\cite{vicentearche_spincharge_2021} and KTO(111)\cite{mallik_superfluid_2022,mallik_electronic_2023}, a redox reaction transforms the Al in AlO$_x$ and induces oxygen vacancies in the KTO, forming the 2DEG. Through an advanced spectro-microscopy analysis at the picometer scale we first report the intimate atomic structure of the KTO(110) 2DEG. We concomitantly observe (i) the elongation of the KTO unit cell, (ii) polar displacements of the ionic species and (iii) a reduction of the Ta valence state, in the 2DEG region located within $\SI{2}{\nano\meter}$ from the interface with the AlO$_x$ overlayer. We then perform band structure calculations using density functional theory and compare them with experimental observations of the band dispersion and highly anisotropic Fermi surfaces using ARPES. Equipped with these results we compute the spin and Edelstein orbital effects and find significant orbital contributions and anisotropy with respect to the crystallographic axes in the plane. We finally probe the Edelstein effect through harmonic transport (nonreciprocal magnetoresistance) and its in-plane anisotropy, and by comparison with theory conclude that it is predominantly determined by the OEE. 

\begin{figure*}[t]
\centering
\includegraphics[width=\textwidth]{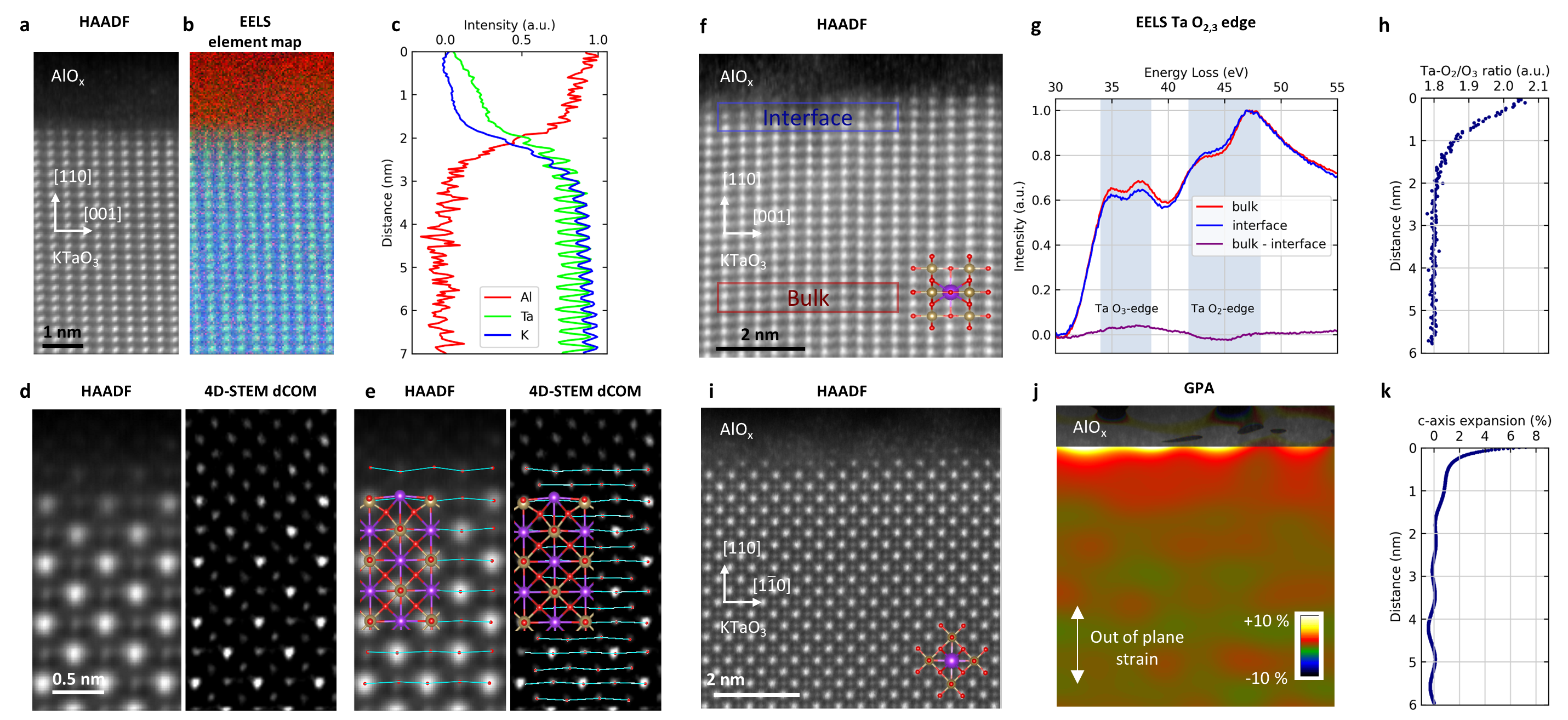}
\caption{\label{1_STEM}\textbf{Atomic structure of the AlO$_x$/KTaO$_3$(110) interface.}
(a) HAADF-STEM image of the AlO$_x$/KTO interface observed along zone axis [$1 \bar{1} 0$]. (b) Corresponding atomically resolved EELS element map from Ta-M$_{4,5}$, K-L$_{2,3}$, and Al-K edges. (c) Intensity profiles of different components in this element map. (d) HAADF-STEM and d-COM images from 4D-STEM. (e) Same images with overlapped DFT relaxed KTO (110) surface and cyan lines connecting atomic positions measured from the experimental images. (f) Large scale HAADF-STEM image along zone axis [$1 \bar{1} 0$]. (g) EELS of the Ta $O_2,_3$ edges at the region near the interface and the bulk as highlighted in the HAADF image in (f). (h) Real space intensity ratio of Ta-$O_2$ and Ta-$O_3$ edges from the different regions in the HAADF image. $O_{2}$/$O_{3}$ ratios were estimated by integrating the two energy loss bands, 34-39 eV and 42-48 eV, shown in (g). (i) HAADF-STEM image of the AlO$_x$/KTO interface observed along zone axis [001], (j) Out-of-plane lattice expansion obtained by GPA, and (k) its profile.}   
\end{figure*}

\section*{Scanning transmission electron microscopy}

High-resolution images of the AlO$_x$/KTaO$_3$~(110) interface were obtained by high-angle annular dark-field (HAADF) scanning transmission electron microscopy (STEM). Fig.~\ref{1_STEM}a,d,f,i show cross-sectional views along two perpendicular zone axes, where the interface between the annealed KTO substrate and the amorphous AlO$_x$ capping layer appears abrupt, with the heavier Ta atoms providing strong contrast in the HAADF images. Electron energy-loss spectroscopy (EELS) (Figs.~\ref{1_STEM}b,c) reveals partial interdiffusion of Ta and K into the AlO$_x$. Within the first three unit cells, an imbalance between Ta and K is observed, consistent with partial K volatilization during annealing. Such deviations from stoichiometry, reported previously, are thought to reduce the polar discontinuity of the alternating [KTaO]$^{4+}$/[O$_2$]$^{4-}$ sublayers\cite{xu_giant_2024}.

Lattice distortions were quantified using geometrical phase analysis (GPA) of HAADF images. A gradual out-of-plane expansion of the KTO lattice appears within $\sim$2~nm of the interface, with the last unit cell showing an increase of $\sim$5\% relative to bulk KTO (Fig.~\ref{1_STEM}j,k). 
Polar displacements of the Ta cations relative to the surrounding K square were also detected, indicating a polar-type distortion of the out-of-plane lattice (Fig.~\ref{1_STEM}d,e). These distortions, visible in both HAADF and 4D-STEM images, reach amplitudes up to $\sim$25~pm in the interfacial unit cell and decay over $\sim$2~nm into the substrate. Consistent with prior reports on perovskite (110) surfaces\cite{annadi_anisotropic_2013,pojani_polarity_1999}, we find that polarity is compensated by rumpling of the Ta–O planes. 
Our density-functional theory (DFT) calculations reproduce this distortion, with Ta ions shifted by $\sim$51~pm. This value is higher than that observed by STEM, likely because the polar discontinuity is already partially reduced at the interface compared to the free surface. Overall, the observed displacements of both cations and oxygen indicate a broken symmetry and polar-type distortion in the interfacial 2DEG region. Finally, the Ta valence was probed by analyzing the fine structure of the Ta $O$-edge in EELS. Layer-resolved spectra (Fig.~\ref{1_STEM}g) show a reduction in Ta valence that saturates $\sim$2~nm from the interface, reaching a maximum decrease of $\sim$12\%, cf. Fig.~\ref{1_STEM}h. This confirms the presence of a reduced Ta state in the 2DEG, in agreement with the structural distortions identified above.

\begin{figure*}[t]
\centering
\includegraphics[width=\textwidth]{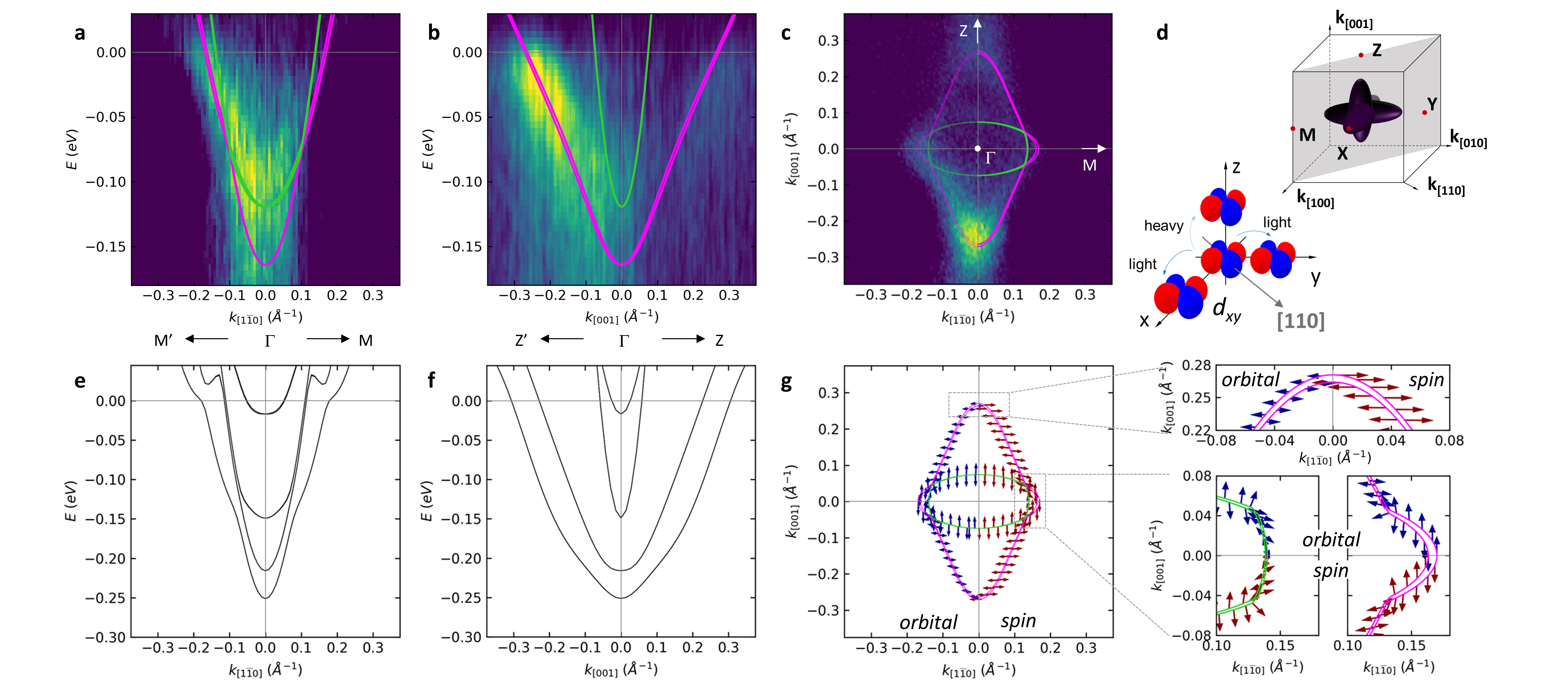}
\caption{\label{arpes}\textbf{Electronic band structure of the KTaO$_3$(110) 2DEG.}
(a,b) ARPES dispersions along $\Gamma$–M ([$1\bar{1}0$]) and $\Gamma$–Z ([$001$]) for EuO$_x$/KTO(110), overlaid with tight-binding fits including spin–orbit coupling.  
(c) Constant-energy map at the Fermi level showing two orthogonal elliptical Fermi pockets.  
(d) Brillouin-zone cut for the (110) orientation with schematic orbital-selective hopping: mixed $d_{xz/yz}$ along $\Gamma$–M and dominant $d_{xy}$ along $\Gamma$–Z.  
(e,f) Band dispersions from DFT using the relaxed interfacial structure extracted from STEM.  
(g) Calculated spin (red) and orbital (blue) angular-momentum textures at the Fermi surface.  
The anisotropic Rashba splittings and orbital-momentum locking revealed here underpin the orbital Edelstein response discussed below.}
\end{figure*}

\section*{Density functional theory calculations}

To connect the observed orbital textures to the underlying atomic structure, we carried out density functional theory (DFT) calculations using the relaxed interfacial structure extracted from STEM. Fig.~\ref{arpes}e,f show the resulting band dispersions of the KTO (110) 2DEG along the high-symmetry directions $\Gamma$–M and $\Gamma$–Z corresponding to the [$1\bar{1}0$] and [$001$] directions, respectively. The calculations yield two light and two heavy subbands 
and reveal a strong anisotropy between the $\Gamma$–M and $\Gamma$–Z directions, with a significantly flatter dispersion along $\Gamma$–Z.

Note that these calculations have been carried out for a KTO (110) surface, because modeling an interface with AlO$_x$ or EuO$_x$ is too challenging. As a result, the precise subband structure and location of the Fermi level differ from the ARPES measurements presented in the following. Still, qualitative agreement can be observed.

\section*{Experimental determination of the 2DEG electronic band structure}

To determine the detailed band structure experimentally, we performed angle-resolved photoemission spectroscopy (ARPES) on KTO~(110)-based two-dimensional electron gases (2DEGs) obtained by $\emph{in situ}$ Eu and Al deposition. The data shown here correspond to 2.4~\AA{} of Eu; similar results were obtained with Al (see Supplementary Information). Fig.~\ref{arpes}a,b display the energy dispersions along $\text{M}'$–$\Gamma$–M and $\text{Z}'$–$\Gamma$–Z. Fig.~\ref{arpes}c presents the constant-energy map at the Fermi level. The data are generally consistent with the DFT results, though some subbands are absent in the ARPES measurement.
The conduction bandwidth is about $-0.17$~eV. The Fermi surface consists of two perpendicular ellipses with anisotropic Fermi wave vectors of 0.17 and 0.27~\AA$^{-1}$ along $\Gamma$–M and $\Gamma$–Z, respectively, revealing strong in-plane anisotropy. The tight-binding fit (TB) reproduces two main pairs of subbands: heavy (pink) and light (green) doublets. The dispersions originate from Ta~5$d$ $t_{2g}$ orbitals, whose degeneracy is lifted by spin-orbit coupling and the (110)-oriented interfacial potential. Along $\Gamma$–M, mixed $d_{xz/yz}$ orbitals form zigzag hopping paths, while along $\Gamma$–Z$\,\,$the $d_{xy}$ character dominates, producing lighter masses. These directional overlaps explain both the elliptic Fermi contours and the anticrossing observed at $-0.08$~eV along $\Gamma$–M. Polarization-dependent ARPES using circular and linear light yielded no strong orbital selectivity, yet comparison with TB and symmetry considerations supports this assignment. The same hierarchy between $d_{xy}$ and $d_{xz/yz}$ states has been reported for KTO~(110) and related orientations~\cite{chen_orientation-dependent_2024,varotto_direct_2022,mallik_electronic_2023}, independent of the overlayer used (EuO$_x$, AlO$_x$, or LaAlO$_3$).

From the TB analysis we estimate a sheet carrier density $n_s\!\approx\!6.8\times10^{13}$~cm$^{-2}$. EuO$_x$/KTO(110) 2DEGs with similar densities are known to be superconducting with $T_c$ around 1 K~\cite{chen_two-dimensional_2021,hua_superconducting_2024}. Our system therefore lies within the superconducting regime, and accordingly the samples are superconducting (at 0.7 K, cf. Supplementary Information). The effective-mass anisotropy is stronger than in STO(110), consistent with the heavier Ta ions and stronger SOC in KTO. For the pink (outer) bands, $m^*\!=\!1.3\,m_e$ and $0.9\,m_e$ along $\Gamma$–M and $\Gamma$–Z$, $ while the green (inner) bands remain lighter, $m^*\!=\!0.83\,m_e$ and $0.17\,m_e$, respectively. At lower energies, the pink bands lighten near the $\Gamma$ point due to the anticrossing, reaching 0.42 and 0.57~$m_e$. The corresponding Rashba splittings are $\Delta k=6.8\times10^{-3}$~\AA$^{-1}$ and $6.7\times10^{-3}$~\AA$^{-1}$, giving $\alpha_R\!\approx\!20$~meV$\cdot$\AA{} for the pink bands. For the green bands, $\Delta k=1.5\times10^{-3}$~\AA$^{-1}$ gives $\alpha_R\!\approx\!6$~meV$\cdot$\AA{} along $\Gamma$–M, while almost vanishing along $\Gamma$–Z. These values agree with TB predictions based on maximally localized Wannier functions~\cite{martinez_anisotropic_2023} and with earlier ARPES studies of KTO~(111)~\cite{mallik_electronic_2023}. 

The spin and orbital angular-momentum textures calculated from the TB model are shown in Fig.~\ref{arpes}g. In both subbands, the angular momenta lie in-plane. 
The inner and outer subbands display opposite helicities; however, the orbital component dominates throughout the Fermi surface and persists where the spin polarization vanishes. This strong orbital–momentum locking, together with the anisotropic Rashba splitting, provides the microscopic basis for the orbital Edelstein effect described below. The textures are reminiscent of those of the KTO(111) 2DEG~\cite{mallik_electronic_2023} but deviate from an isotropic Rashba model. 

\section*{Boltzmann - Edelstein effect}

\begin{figure*}[t]
\centering
\includegraphics[width=\textwidth]{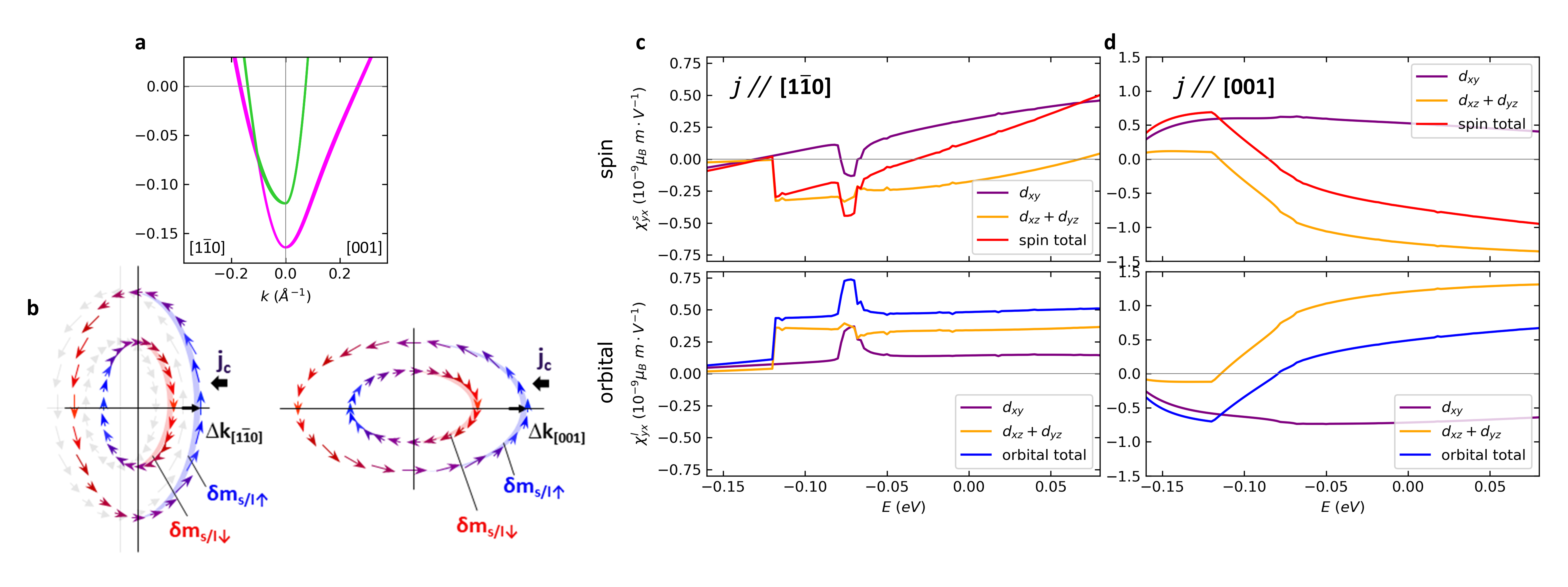}
\caption{\label{boltzmann}\textbf{Calculated spin and orbital Edelstein responses.}
(a) Band structure of the tight-binding model along $\Gamma$–M ([$1\bar{1}0$]) and $\Gamma$–Z ([$001$]) directions, including spin–orbit coupling and confinement potential.  
(b) Schematic representation of the uncompensated densities of spin and/or orbital moments under application of a charge current through the Rashba-Edelstein effect.
(c,d) Energy-dependent Edelstein susceptibility tensor components $\chi^{s}_{ij}$ (spin) and $\chi^{l}_{ij}$ (orbital) computed within the Boltzmann formalism at $T=0$ K for $\tau_0=1$ ps.  
In the [$001$] direction, the orbital response (blue) has an opposite sign to the spin response (red)  across the entire energy range, and in the [$1\bar{1}0$] direction, the two contribution compensate at low energy and add up close to and above the Fermi level.  
}
\end{figure*}

Armed with the TB fit of the experimental band structure, we now calculate the spin and orbital Edelstein effects for the KTO (110) system. An electric field $\mathbf E$ applied to the 2DEG induces a magnetic moment $\mathbf m_0$ per unit cell due to the spin and orbital Edelstein effect, as sketched in Fig.~\ref{boltzmann}a. Following Ref.~\cite{johansson_spin_2021}, we define the linear Edelstein susceptibility tensor $\chi$ as
\begin{align}\label{eq:Edelstein_efficiency}
    \mathbf m_0 = \chi \mathbf E = \left( \chi^\text s + \chi^\text l \right) \mathbf E \ .
\end{align}
Solving the linearized Boltzmann equation within constant relaxation time approximation, as introduced in detail in Ref.~\cite{johansson_spin_2021}, and assuming $T=0$, the spin and orbital susceptibilites are given by
\begin{align}\label{eq:spin_orb_susceptibility}
    \chi^{\text{s/l}}_{ij}= \frac{e g_\text{s/l} \mu_\text B \tau_0 A_0}{\hbar A} \sum \limits_{\mathbf k, n} s/l_{\mathbf k, n} ^i v_{\mathbf k,n} ^j \delta \left( E_{\mathbf k,n} - E_\text F \right) 
\end{align}
with $e$ the absolute value of the elementary charge, $g_\text{s(l)}$ the spin (orbital) Landé $g$ factor, $\mu_\text B$ the Bohr magneton, $\hbar$ the reduced Planck constant, $\tau_0$ the momentum relaxation time,  $A_0$ the area of the unit cell, $A$ the area of the sample, $\mathbf k$ the crystal momentum, $n$ the band index, $\mathbf s_{\mathbf k, n}$ the $\mathbf k$ and band dependent spin expectation value, $\mathbf l_{\mathbf k, n}$ the orbital angular momentum expectation value, $\mathbf v_{\mathbf k, n}$ the group velocity, $E_{\mathbf k, n}$ the energy of a state, $E_\text F$ the Fermi energy, and $i,j$ Cartesian coordinates. Fig.~\ref{boltzmann}c and d show the Edelstein susceptibility tensor elements for KTO~(110) 2DEGs, representing the magnetic moment induced by an electric field applied in [$001$] and [$1 \bar{1} 0$] direction, respectively. The tensor elements are resolved on the different orbitals of the TB model across the band structure. Due to the reduced symmetry of the 2DEG the Edelstein response is anisotropic. 

An electric field applied in [$1 \bar{1} 0$] direction induces a total Edelstein magnetic moment oriented along [$001$] due to the addition of spin- and orbital responses at the Fermi level. Remarkably, the orbital contribution remains constant over a large energy range, whereas the spin Edelstein effect exhibits a linear energy dependence and even a sign change around -$\SI{40}{\milli\electronvolt}$. Hence, the total Edelstein susceptibility $\chi_{xy}$ increases almost linearly in energy without any sign change in the energy region of interest, above that threshold and across the Fermi level. 
The crossing of the band pairs around -$\SI{70}{\milli\electronvolt}$ manifests as a peak in the Edelstein signal.

An electric field applied along the [$001$] direction induces a magnetic moment (total Edelstein effect) along [$\bar{1}10$] for $E_F$ > -$\SI{80}{\milli\electronvolt}$, due to the partial compensation of the spin response by the orbital response. Notably, the spin and orbital Edelstein effects have an opposite sign in the whole energy range, both exhibiting a sign change around -$\SI{80}{\milli\electronvolt}$. 

Our calculations demonstrate that both the magnitude and orientation of the current-induced magnetic moment depend on the direction of the applied charge current. This holds for the individual spin and orbital contributions but also for the total Edelstein signal as they are combined.

\section*{Bilinear Magnetoresistance}

We now probe the Edelstein effect through the current-induced generation of spin and orbital angular momentum densities, using angle-dependent harmonic magnetotransport. In a Rashba system, an electric current produces an interfacial magnetization, which can add or compete with an external magnetic torque depending on their relative orientation. The resulting unidirectional or bilinear magnetoresistance (BMR) thus reflects the interplay between current-induced spin–orbital polarization and the applied magnetic field~\cite{vaz_determining_2020,guillet_observation_2020}.

\begin{figure*}[h!]
\centering
\includegraphics[width=1.0\textwidth]{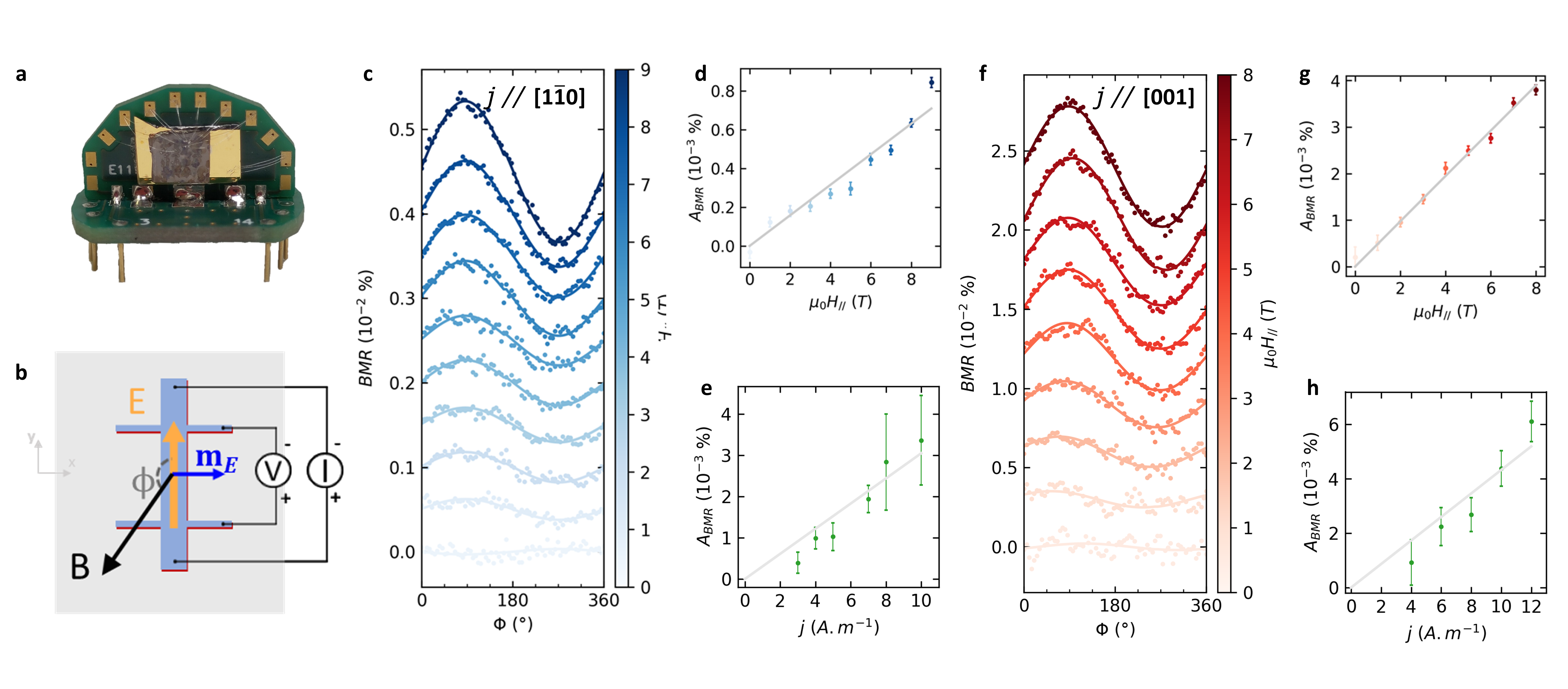}
\caption{\label{BMR} \textbf{Magnetoresistance measurement of the Edelstein effect in a KTO(110)-based 2DEG.}
(a) Photograph of a Hall-bar device used for angular-dependent harmonic magnetotransport.
(b) Schematic of the measurement configuration: the current-induced moment $\mathbf{m}_E$ interacts with an external magnetic field $\mathbf{B}$, producing an angular-dependent modulation of the second-harmonic longitudinal resistance $R^{2f}_{xx}$.
(c,f) BMR signals for current along [$1\bar{1}0$] and [$001$] at different magnetic fields; (d,g) dependence of the BMR amplitude $A_{\text{BMR}}$ on the external magnetic field; (e,h) dependence on the current amplitude.
All measurements were performed at 2~K on AlO$_x$/KTO(110), with $j=$4~A/m for [$1\bar{1}0$] and 10~A/m for [$001$].
The grey lines are linear fits.}
\end{figure*}

Applying an AC current bias, the angular modulation of the second harmonic magnetoresistance takes the form
\begin{align}\label{eq:BMR}
    R^{2f}_{xx} = A_{\text{BMR}}\sin(\phi),
\qquad
A_{\text{BMR}} = \frac{3\pi}{2}\frac{g\mu_B\alpha_R\tau^2}{|e|\hbar \varepsilon_F}\,jB,
\end{align}
where $\phi$ is the in-plane angle between current and field, $j$ is the current density, $B$ the magnetic field, $\mu_B$ the Bohr magneton and $\tau$ the relaxation time (Fig.\ref{BMR}b). The BMR amplitude therefore scales linearly with both $j$ and $B$, as observed experimentally.

Hall bars patterned along the [$001$] and [$1\bar{1}0$] directions of AlO$_x$/KTO(110) were measured at 2~K in the normal state. The first and second harmonics of the longitudinal resistance were recorded by lock-in detection while rotating the magnetic field within the film plane. Fig.\ref{BMR}c and \ref{BMR}f show representative $R^{2f}_{xx}(\phi)$ traces for increasing $B$ and $j$ along both current directions. Data were antisymmetrized as $R_{xx}(2f)(\phi)=\big[R_{xx}(B,+\phi)-R_{xx}(B,-\phi)\big]/2$ to remove residual thermal and misalignment effects.

The second-harmonic response follows a clear sinusoidal dependence, confirming the unidirectional character of the signal. The BMR amplitude $A_{\text{BMR}}$ increases linearly with both $j$ and $B$ (Fig.\ref{BMR}d,e,g,h), consistent with Eq.~(3) and characteristic of an Edelstein mechanism.
The experimental BMR signals exhibit the same sign for both current directions. It is negative, as in KTO(001) \cite{vicentearche_spincharge_2021} and opposite to that in STO(001) 2DEGs \cite{vaz_determining_2020}, indicating here a globally negative Edelstein effect. 
As can be seen in the measurement configuration displayed in Fig.\ref{BMR}b, the negative sign of the BMR indicates that for both directions, the magnetic moment induced by the Edelstein effect is always at +90° relative to the electric field. Indeed, the high resistance state is obtained at a rotation angle of 90°, that is when the external magnetic field interferes constructively with the pseudo-field of the current-induced magnetic moment.
This is consistent with the signs predicted by the Edelstein tensor calculations from Boltzmann transport, with the chosen convention for the in-plane directions. 

To compare the Edelstein efficiencies between the directions, we define the normalized quantity 
$\eta = \frac{A_{\text{BMR}}}{jB},$
and obtain $\eta_{1\bar{1}0} \approx 1.0\times10^{-11}$~cm$^2$/A$\cdot$T and $\eta_{001} \approx 2.5\times10^{-11}$~cm$^2$/A$\cdot$T, values comparable to single-layer Bi$_2$Se$_3$~\cite{he_bilinear_2018}, and orders of magnitude higher than STO- or LAO/KTO-based interfaces~\cite{chen_two-dimensional_2021,liu_tunable_2023}. 
With the identical sign and a magnitude ratio of about 2.5, the bilinear magnetoresistance response is indeed anisotropic, in line with expectations drawn from the band structure. 

Effective Rashba parameters can also be extracted from Eq.~(3). With a carrier density of $5.8\times10^{13}$~cm$^{-2}$ measured by Hall effect, the scattering times are estimated from the normal state resistivity in the Drude model as
$\tau\simeq73$~fs for [$1\bar{1}0$] with the weighted effective mass $m^*_{1\bar{1}0}=1.08~ m_e$ and 419~fs for [$001$] with $m^*_{001}=3.7~ m_e$. Using $\varepsilon_F\!=\!170$~meV and a prefactor $g$ between 0.5 and 2, we find effective Rashba coefficients $\alpha_R$ of -9 – -74 for [$1\bar{1}0$] and and -0.9 – -3.9~meV$\cdot$\AA{} for [$001$]. These values are indeed well anisotropic, and are comparable or slightly larger than those deduced from the band structure and comparable to recent reports on LAO/KTO(110)~\cite{liu_tunable_2023}.

Coming back to the Boltzmann calculations of Fig.~\ref{boltzmann}c,d, the signs and magnitude ratio of experimental data thus suggest that the orbital Edelstein effect strongly contributes to the experimental BMR results. Indeed, in the region of interest, between the Fermi level and the binding energy of $-0.0185$ eV corresponding to the carrier density of the measured sample in the TB model, neither spin nor orbital Edelstein effect alone explain the experimental results. 
If the experimentally determined signs match that of the spin contribution to the Edelstein effect, the magnitude of the effects are only reproduced by accounting for the orbital contribution in a proportion of 10–30\% of the total Edelstein effect.

\section*{Conclusion}

We have established contributions of orbital Edelstein effect in KTaO$_3$(110) two-dimensional electron gases by closing the loop from atomic structure to nonreciprocal transport. Aberration-corrected STEM/EELS reveals interfacial rumpling and a reduced Ta valence within $\sim$2\,nm of the surface, which we use to construct a realistic structural model for density-functional calculations. The resulting band structure – anisotropic elliptical Fermi surfaces with a hybridized $t_{2g}$ manifold – is confirmed by ARPES and captured by a tight-binding model that includes the interfacial potential and spin-orbit coupling. Boltzmann calculations based on this model predict an anisotropic Edelstein response in which the orbital contribution adds to the spin contribution over a broad energy window and fixes the anisotropy.

Angle-dependent bilinear magnetoresistance measurements validate these predictions. The second-harmonic signals scale linearly with current and field, are negative, and exhibit the same sign for currents applied along [$1\bar{1}0$] and [$001$]. Together with the calculated anisotropy, these observations identify a current-induced \emph{orbital} polarization as a necessary source of the measured nonreciprocal transport and demonstrate the presence of the orbital Edelstein effect in KTaO$_3$(110).

Beyond establishing orbital Edelstein physics in oxide 2DEGs, our results position KTaO$_3$(110) as a benchmark system for orbitronics, where current-induced phenomena feature orbital as well as spin textures. The same orbital degrees of freedom that amplify the Edelstein response may also couple to superconducting pairing, offering a new perspective on the origin and tunability of superconductivity in KTaO$_3$-based interfaces. Exploring how orbital polarization evolves under gate control, strain, or orientation could thus reveal pathways toward coherent orbital manipulation and dissipationless orbital transport in complex oxides.

\section*{Acknowledgements}

This project received funding from the European Research Council Advanced Grant “FRESCO” ($\#$833973), the QuantERA project "QUANTOX" (ANR-18-QUAN-0014), the French ANR projects "QUANTOP" (ANR-19-CE47-0006-01), "SURIKAT" (ANR-23-CE30-0036), "ImagingQM" (ANR-23-CE42-0027) and the European Innovation Council Horizon Europe projects "IQARO" ($\#$101115190) and "OBELIX" ($\#$101129641). This work was funded by the Deutsche Forschungsgemeinschaft (DFG, German Research Foundation) – 545818886.

\section*{Author contributions statement}

Samples were prepared by H.W., S.M., L.M.V.A. and S.V. ARPES measurements were performed by S.M., L.M.V.A., J.B. and S.V. with help from R.S., J.R., F.B. and P.L.F. The ARPES data were analyzed by S.M., S.V. and H.W. STEM, EELS and DFT calculations were performed by A.R. and A.G. TB fits and Edelstein calculations were performed by B.G., A.J. and I.M. Transport measurements were performed and data analyzed by H.W. with help from J.B., G.M., A.B., N.B. and M.B. The manuscript was written by H.W., M.B., A.G., B.G. and A.J. with inputs from all authors. The project was conceived and supervised by M.B. with N.B., A.B., I.B., A.J. and I.M.

\section*{Additional information}

The authors declare no competing interests. 

\section*{Methods}

\subsection*{Sample preparation}
For TEM, the sample was made by magnetron sputtering.
After ultrasonic cleaning in EDI, acetone and isopropanol, and in-situ pre-annealing at 600 $\degree$C for 30 minutes, 2.1 nm of Al was sputtered with a DC current of 30 mA, base pressure of Ar of 4.10$^{-3}$ mbar. The deposition was done in two steps, first at 500$\degree$C and second at a temperature below 50$\degree$C.

For transport measurement, the Hall bar devices were patterned using a hard mask template method, as described in \cite{witt_patterning_2023}. Using standard optical lithography and lift-off process, a 50 nm thick insulating Al$_2$O$_3$ film was deposited on the surface by RF sputtering off an Al$_2$O$_3$ target, patterning 100 nm long, 10 and 25 nm wide bars.
The deposition of Al for the creation of the 2DEG was carried out by magnetron sputtering with the same recipe as for the sample for TEM, although with a deposition temperature of the first step at 450$\degree$C.

For ARPES measurements, the sample was made by molecular beam epitaxy, in the same way as in \cite{mallik_electronic_2023}. The substrate was pre-annealed at 300$\degree$C for 2 hours in vacuum, then 2.4~\AA of Eu were grown at 300$\degree$C with a growth rate of 0.025 $\text{\AA}$.s$^{-1}$ in a deposition pressure of 1.7~$\times$ 10$^{-8}$ mbar.

Commercial single crystalline (110)-oriented KTaO$_3$ substrates were supplied by MTI Corporation.

\subsection*{STEM and EELS}

The cross-sectional lamellae for STEM analysis were prepared using a Focused Ion Beam (FIB) technique. Prior to FIB milling, a protective layer of approximately 30 nm of amorphous carbon was deposited on the sample surface. Additionally, electron beam-induced deposition of platinum was performed, followed by ion beam-induced deposition of Pt, both serving as protective layers during the milling process.

STEM experiments were conducted using a Cs-corrected NION UltraSTEM and a Cs-corrected, monochromated NION Chromatem microscope. 4D-STEM and STEM-EELS elemental mapping were performed at an accelerating voltage of 200 keV on the UltraSTEM, with a probe current of approximately 10 pA and a convergence semi-angle of 30 mrad. The divergences of the in-plane electric field (estimated from the negative center-of-mass, or -COM, signals) were also computed. These divergence maps (dCOM), which approximate the projected charge-density distribution, show extrema at atomic positions and enable analysis of oxygen positions that are not visible in HAADF images. Corrugation was estimated by measuring the atomic positions using a Gaussian fitting, as implemented in the Atomap software\cite{Nord2017}. STEM-EELS elemental maps were obtained by simultaneously collecting the Al--K, Ta--M$_{4,5}$, and K--L$_{2,3}$ edges.

Monochromated EELS measurements targeting the Ta \( \mathrm{O}_{2,3} \) edges were conducted at 100 keV using the NION Chromatem. These measurements were carried out with a probe current of 10 pA and an energy resolution of 120 meV, while maintaining atomic resolution.

\subsection*{Density functional theory}
DFT calculations are performed with OpenMX package\cite{ozaki2003variationally}, with PBE-GGA \cite{perdew1996generalized}
exchange correlation functional. In OpenMX, pseudo atomic orbital method is implemented, with pre-defined pseudo-atomic orbitals provided in the package. A s3p2d1, s3p2d2f1, and s2p2d1 type localized basis sets are used for K, Ta and O atoms respectively. The letters s,p, and d represent the angular dependence of the orbitals, giving 1, 3 and 5 harmonics respectively. This means, 14 localized orbitals for K, 26 localized orbitals for Ta, and 13 localized orbitals for O. A 33-atom superlattice of KTO with the [110] cut surface was considered for the geometrical optimization, The in-plane parameters of it being $\sqrt{2}a$ and $a$ and out-of-plane parameter $\sqrt{2}a$, with $a$ being the lattice parameter of simple cubic KTO. It was carried out with a force cutoff of $3 \times 10^{-4}$ Hartree/Bohr. The spin polarized calculation and bandstructure calculations were carried out in the optimized structure with non-collinear spin orientation.

\subsection*{Angle resolved photoemission spectroscopy}
High-resolution angular resolved photoemission spectroscopy (ARPES) measurements were conducted at the Cassiopée beamline of Synchrotron SOLEIL (France), using a Scienta R4000 hemispherical electron energy analyser. The sample was cooled to 15 K with liquid He during measurements. The energy resolution is 15 meV and the angular resolution is $<$ 0.25$\degree$. Fig \ref{arpes} shows data collected with circular left (CL) polarized photons at a photon energy of 31 eV. The collected data were normalized by taking the second derivative of the intensity background of the electron analyzer and smoothed using an averaging filter.

\subsection*{Tight-binding model}
We consider a tight-binding (TB) model of 24 bands (12 band pairs). One unit cell includes two of each of the 5\textit{d} Ta $t_{2g}$ orbitals ($d_{xy}$, $d_{xz}$, $d_{yz}$), hence the basis 
\begin{align}\label{eq:basis}
\{ \ket{d_{xy1\uparrow}},\ket{d_{xy1\downarrow}} ,\ket{d_{yz1\uparrow}},\ket{d_{yz1\downarrow}} ,\ket{d_{zx1\uparrow}},\ket{d_{zx1\downarrow}},\ket{d_{xy2\uparrow}},\ket{d_{xy2\downarrow}} ,\ket{d_{yz2\uparrow}},\ket{d_{yz2\downarrow}} ,\ket{d_{zx2\uparrow}},\ket{d_{zx2\downarrow}}&\}.
\end{align}
The subbands emerging from the confinement of electrons at the interface are accounted for by the addition of a copy of the bands with different parameters.
We diagonalize the tight-binding Hamiltonian
\begin{align}\label{eq:Hamiltonian}
H=H_\mathrm{hop1}+H_\mathrm{hop2}+H_\mathrm{mix}+H_\mathrm{SOC}
\end{align}
to describe the 2DEG at the interface of KTO~(110), with a lattice constant $a=4.0~\,\mathrm{\mathring{A}}$.

For the first and second nearest neighbor hopping, we consider 
%
\begin{align}
H_\mathrm{hop1}=&\begin{psmallmatrix}
2t_\delta ~\mathrm{cos}(k_z a) + \Delta \epsilon_1 &0&0&t_{\pi}h_{-x} + t_{\pi}h_{-y}&0&0\\
0&2t_\pi ~\mathrm{cos}(k_z a) + \Delta \epsilon_2 &0&0&t_{\delta}h_{-x} + t_{\pi}h_{-y}&0&\\
0&0&2t_\pi ~\mathrm{cos}(k_z a) + \Delta \epsilon_2     &0&0&t_{\pi}h_{-x} + t_{\delta}h_{-y}\\
t_{\pi}h_x + t_{\pi}h_y&0&0&2t_\delta ~\mathrm{cos}(k_z a) + \Delta \epsilon_1 &0&0\\
0&t_{\delta}h_x + t_{\pi}h_y&0&0&2t_\pi ~\mathrm{cos}(k_z a) + \Delta \epsilon_2 &0\\
0&0&t_{\pi}h_x + t_{\delta}h_y&0&0&2t_\pi ~\mathrm{cos}(k_z a) + \Delta \epsilon_2\\
\end{psmallmatrix}
\bigotimes
\begin{pmatrix} 1&0\\0&1 \end{pmatrix}
\end{align}

\begin{align}
H_\mathrm{hop2}=&\begin{psmallmatrix}
0&0&0&0&0&0\\   
0&2(t_{2\pi} + t_{2\delta}) ~\mathrm{cos}(ak_x - ak_y)&0&0&0&0\\
0&0&2(t_{2\pi} + t_{2\delta}) ~\mathrm{cos}(ak_x - ak_y)&0&0&0\\
0&0&0&0&0&0\\
0&0&0&0&2(t_{2\pi} + t_{2\delta}) ~\mathrm{cos}(ak_x - ak_y)&0\\
0&0&0&0&0&2(t_{2\pi} + t_{2\delta}) ~\mathrm{cos}(ak_x - ak_y)\\
\end{psmallmatrix}
\bigotimes
\begin{pmatrix} 1&0\\0&1 \end{pmatrix}
\end{align}

with $h_{\pm j} = e^{\pm ik_ja}$ and $\alpha=\{\pi,\delta\}$ indicating the nearest-neighbor hopping of the $\{d_{xy},d_{yz},d_{zx}\}$ orbitals along $\{x,y,z\}$ according to the Slater-Koster rules. The linearly independent hopping amplitudes are $t_{\pi}=-1.37~\,\mathrm{eV}$, $t_{\delta}=-0.07~\,\mathrm{eV}$. Note, that the $\sigma$ hopping is not relevant for nearest-neighbor hoppings considering only the $t_{2g}$ orbitals. The onsite energies are $\Delta \epsilon_1=2.83~\,\mathrm{eV}$ for the $d_{xy}$ orbitals and $\Delta \epsilon_2=4.39~\,\mathrm{eV}$ for the $d_{yz}$ and $d_{zx}$ orbitals causing a shift of the bands in energy.
In the second nearest neighbor hopping, along [1$\overline{1}$0] and [$\overline{1}$10], only the diagonal terms indicating hopping between two $d_{yz}$ orbitals and between two $d_{zx}$ orbitals are relevant, with a hopping amplitude $t_{2\pi} + t_{2\delta}=-98~\,\mathrm{meV}$.

The gradient potential at the interface displaces the oxygens' $p$ orbitals away from the bond connecting two neighboring Ta atoms. Hopping terms that are forbidden in bulk are now allowed by the Slater-Koster rules. By analogy with KTO~(001) 2DEGs, discussed in Ref.~\cite{varotto_direct_2022}, hopping between two neighboring Ta $d$ orbitals, via an intermediate hopping to an oxygen $p$ orbital, causes an effective hopping network. The hopping is asymmetric and gives rise to orbital mixing terms
%
\begin{align}
H_\mathrm{mix}=&g\begin{psmallmatrix}
0&-2i~\mathrm{sin}(k_za)&-2i~\mathrm{sin}(k_za)&0&0&0\\
2i~\mathrm{sin}(k_za)&0&0&0&0&-h_{-y}+h_{-x}\\
2i~\mathrm{sin}(k_za)&0&0&0&h_{-y}-h_{-x}&0\\
0&0&0&0&-2i~\mathrm{sin}(k_za)&-2i~\mathrm{sin}(k_za)\\
0&0&h_{y}-h_{x}&2i~\mathrm{sin}(k_za)&0&0\\
0&-h_{y}+h_{x}&0&2i~\mathrm{sin}(k_za)&0&0
\end{psmallmatrix}
\bigotimes
\begin{pmatrix} 1&0\\0&1 \end{pmatrix}
\end{align}
with $h_{\pm j} = e^{\pm ik_ja}$, $g=5~\,\mathrm{meV}$.

$H_\mathrm{SOC}$ with $\lambda=0.4~\,\mathrm{eV}$ describes the on-site spin-orbit coupling which mixes the different spin orientations and orbitals. It is the same for each lattice site and reads
\begin{align}
H_\mathrm{SOC}=\frac{\lambda}{3}\begin{pmatrix}
0&0&0&0&1&-i\\
0&0&i&-1&0&0\\
0&-i&0&i&0&0\\
0&-1&-i&0&0&0\\
1&0&0&0&0&-i\\
i&0&0&0&i&0
\end{pmatrix}
\end{align}
in the basis $\{ \ket{d_{xy\uparrow}},\ket{d_{xy\downarrow}} ,\ket{d_{yz\uparrow}},\ket{d_{yz\downarrow}} ,\ket{d_{zx\uparrow}},\ket{d_{zx\downarrow}} \}$.

\subsection*{Magnetotransport measurements}
Electrical contacts to the 2DEG were made by ultrasonic wedge bonding with Al wires on Hall bar devices patterned along the [001] and [1$\overline{1}$0] crystallographic direction. AC magnetotransport measurements were conducted with a PPMS Dynacool system by Quantum Design. Angle-dependent measurement are performed by rotating the sample in the same plane as the magnetic field inside the cryostat, with constant values of temperature, magnetic field and current. By convention, angle $\phi = 0 \degree$ is when the magnetic field and current are in the same direction. AC current at a frequency of 148.27 Hz  was applied to the Hall bars with a Keithley 6221 current source and the first and second harmonic of the longitudinal and transverse voltage signals were extracted with a Zurich Instruments HF2LI lock-in amplifier. Fig \ref{BMR} shows the 2f imaginary part of the longitudinal signal. To account for residual out-of-plane magnetoresistance, the data is antisymmetrized with respect to the magnetic field and normalized with a slight linear slope. Relative values as on the ordinate axis correspond to the angle dependent modulation relative to the sheet resistance of the device at angle 0, and curves are shifted by an additional arbitrary offset.


\begin{thebibliography}{10}
\urlstyle{rm}
\expandafter\ifx\csname url\endcsname\relax
  \def\url#1{\texttt{#1}}\fi
\expandafter\ifx\csname urlprefix\endcsname\relax\def\urlprefix{URL }\fi
\expandafter\ifx\csname doiprefix\endcsname\relax\def\doiprefix{DOI: }\fi
\providecommand{\bibinfo}[2]{#2}
\providecommand{\eprint}[2][]{\url{#2}}

\bibitem{hwang_emergent_2012}
\bibinfo{author}{Hwang, H.~Y.} \emph{et~al.}
\newblock \bibinfo{journal}{\bibinfo{title}{Emergent phenomena at oxide interfaces}}.
\newblock {\emph{\JournalTitle{Nature Mater}}} \textbf{\bibinfo{volume}{11}}, \bibinfo{pages}{103--113}, \doiprefix\url{10.1038/nmat3223} (\bibinfo{year}{2012}).

\bibitem{reyren_superconducting_2007}
\bibinfo{author}{Reyren, N.} \emph{et~al.}
\newblock \bibinfo{journal}{\bibinfo{title}{Superconducting {Interfaces} {Between} {Insulating} {Oxides}}}.
\newblock {\emph{\JournalTitle{Science}}} \textbf{\bibinfo{volume}{317}}, \bibinfo{pages}{1196--1199} (\bibinfo{year}{2007}).

\bibitem{caviglia_tunable_2010}
\bibinfo{author}{Caviglia, A.~D.} \emph{et~al.}
\newblock \bibinfo{journal}{\bibinfo{title}{Tunable {Rashba} {Spin}-{Orbit} {Interaction} at {Oxide} {Interfaces}}}.
\newblock {\emph{\JournalTitle{Phys. Rev. Lett.}}} \textbf{\bibinfo{volume}{104}}, \bibinfo{pages}{126803}, \doiprefix\url{10.1103/PhysRevLett.104.126803} (\bibinfo{year}{2010}).
\newblock \bibinfo{note}{Publisher: American Physical Society}.

\bibitem{vaz_determining_2020}
\bibinfo{author}{Vaz, D.~C.} \emph{et~al.}
\newblock \bibinfo{journal}{\bibinfo{title}{Determining the {Rashba} parameter from the bilinear magnetoresistance response in a two-dimensional electron gas}}.
\newblock {\emph{\JournalTitle{Phys. Rev. Mater.}}} \textbf{\bibinfo{volume}{4}}, \bibinfo{pages}{071001}, \doiprefix\url{10.1103/PhysRevMaterials.4.071001} (\bibinfo{year}{2020}).
\newblock \bibinfo{note}{Publisher: American Physical Society}.

\bibitem{lesne_highly_2016}
\bibinfo{author}{Lesne, E.} \emph{et~al.}
\newblock \bibinfo{journal}{\bibinfo{title}{Highly efficient and tunable spin-to-charge conversion through {Rashba} coupling at oxide interfaces}}.
\newblock {\emph{\JournalTitle{Nature Mater.}}} \textbf{\bibinfo{volume}{15}}, \bibinfo{pages}{1261} (\bibinfo{year}{2016}).

\bibitem{vaz_mapping_2019}
\bibinfo{author}{Vaz, D.~C.} \emph{et~al.}
\newblock \bibinfo{journal}{\bibinfo{title}{Mapping spin–charge conversion to the band structure in a topological oxide two-dimensional electron gas}}.
\newblock {\emph{\JournalTitle{Nature Mater.}}} \textbf{\bibinfo{volume}{18}}, \bibinfo{pages}{1187--1193}, \doiprefix\url{10.1038/s41563-019-0467-4} (\bibinfo{year}{2019}).
\newblock \bibinfo{note}{Number: 11 Publisher: Nature Publishing Group}.

\bibitem{levitov_magnetoelectric_1985}
\bibinfo{author}{Levitov, L.~S.} \& \bibinfo{author}{Nazarov, Y.~V.}
\newblock \bibinfo{journal}{\bibinfo{title}{Magnetoelectric effects in conductors with mirror isomer symmetry}}.
\newblock {\emph{\JournalTitle{Sov. Phys. JETP}}} \textbf{\bibinfo{volume}{61}}, \bibinfo{pages}{133} (\bibinfo{year}{1985}).

\bibitem{johansson_theory_2024}
\bibinfo{author}{Johansson, A.}
\newblock \bibinfo{journal}{\bibinfo{title}{Theory of spin and orbital {Edelstein} effects}}.
\newblock {\emph{\JournalTitle{J. Phys.: Condens. Matter}}} \textbf{\bibinfo{volume}{36}}, \bibinfo{pages}{423002}, \doiprefix\url{10.1088/1361-648X/ad5e2b} (\bibinfo{year}{2024}).

\bibitem{ding_harnessing_2020}
\bibinfo{author}{Ding, S.} \emph{et~al.}
\newblock \bibinfo{journal}{\bibinfo{title}{Harnessing {Orbital}-to-{Spin} {Conversion} of {Interfacial} {Orbital} {Currents} for {Efficient} {Spin}-{Orbit} {Torques}}}.
\newblock {\emph{\JournalTitle{Phys. Rev. Lett.}}} \textbf{\bibinfo{volume}{125}}, \bibinfo{pages}{177201}, \doiprefix\url{10.1103/PhysRevLett.125.177201} (\bibinfo{year}{2020}).

\bibitem{krishnia_quantifying_2024}
\bibinfo{author}{Krishnia, S.} \emph{et~al.}
\newblock \bibinfo{journal}{\bibinfo{title}{Quantifying the large contribution from orbital {Rashba}–{Edelstein} effect to the effective damping-like torque on magnetization}}.
\newblock {\emph{\JournalTitle{APL Materials}}} \textbf{\bibinfo{volume}{12}}, \bibinfo{pages}{051105}, \doiprefix\url{10.1063/5.0198970} (\bibinfo{year}{2024}).

\bibitem{el_hamdi_observation_2023}
\bibinfo{author}{El~Hamdi, A.} \emph{et~al.}
\newblock \bibinfo{journal}{\bibinfo{title}{Observation of the orbital inverse {Rashba}–{Edelstein} effect}}.
\newblock {\emph{\JournalTitle{Nat. Phys.}}} \doiprefix\url{10.1038/s41567-023-02121-4} (\bibinfo{year}{2023}).

\bibitem{wemple_transport_1965}
\bibinfo{author}{Wemple, S.~H.}
\newblock \bibinfo{journal}{\bibinfo{title}{Some {Transport} {Properties} of {Oxygen}-{Deficient} {Single}-{Crystal} {Potassium} {Tantalate} ({KTa}\$\{{\textbackslash}mathrm\{{O}\}\}\_\{3\}\$)}}.
\newblock {\emph{\JournalTitle{Phys. Rev.}}} \textbf{\bibinfo{volume}{137}}, \bibinfo{pages}{A1575--A1582}, \doiprefix\url{10.1103/PhysRev.137.A1575} (\bibinfo{year}{1965}).
\newblock \bibinfo{note}{Publisher: American Physical Society}.

\bibitem{thompson_very_1982}
\bibinfo{author}{Thompson, J.~R.}, \bibinfo{author}{Boatner, L.~A.} \& \bibinfo{author}{Thomson, J.~O.}
\newblock \bibinfo{journal}{\bibinfo{title}{Very low-temperature search for superconductivity in semiconducting {KTaO3}}}.
\newblock {\emph{\JournalTitle{J Low Temp Phys}}} \textbf{\bibinfo{volume}{47}}, \bibinfo{pages}{467--475}, \doiprefix\url{10.1007/BF00683987} (\bibinfo{year}{1982}).

\bibitem{ueno_discovery_2011}
\bibinfo{author}{Ueno, K.} \emph{et~al.}
\newblock \bibinfo{journal}{\bibinfo{title}{Discovery of superconductivity in {KTaO3} by electrostatic carrier doping}}.
\newblock {\emph{\JournalTitle{Nature Nanotech}}} \textbf{\bibinfo{volume}{6}}, \bibinfo{pages}{408--412}, \doiprefix\url{10.1038/nnano.2011.78} (\bibinfo{year}{2011}).

\bibitem{varotto_direct_2022}
\bibinfo{author}{Varotto, S.} \emph{et~al.}
\newblock \bibinfo{journal}{\bibinfo{title}{Direct visualization of {Rashba}-split bands and spin/orbital-charge interconversion at {KTaO3} interfaces}}.
\newblock {\emph{\JournalTitle{Nature Commun.}}} \textbf{\bibinfo{volume}{13}}, \bibinfo{pages}{6165}, \doiprefix\url{10.1038/s41467-022-33621-1} (\bibinfo{year}{2022}).
\newblock \bibinfo{note}{Number: 1 Publisher: Nature Publishing Group}.

\bibitem{vicentearche_spincharge_2021}
\bibinfo{author}{Vicente‐Arche, L.~M.} \emph{et~al.}
\newblock \bibinfo{journal}{\bibinfo{title}{Spin–{Charge} {Interconversion} in {KTaO} $_{\textrm{3}}$ {2D} {Electron} {Gases}}}.
\newblock {\emph{\JournalTitle{Adv. Mater.}}} \textbf{\bibinfo{volume}{33}}, \bibinfo{pages}{2102102} (\bibinfo{year}{2021}).

\bibitem{liu_two-dimensional_2021}
\bibinfo{author}{Liu, C.} \emph{et~al.}
\newblock \bibinfo{journal}{\bibinfo{title}{Two-dimensional superconductivity and anisotropic transport at {KTaO} $_{\textrm{3}}$ (111) interfaces}}.
\newblock {\emph{\JournalTitle{Science}}} \textbf{\bibinfo{volume}{371}}, \bibinfo{pages}{716--721}, \doiprefix\url{10.1126/science.aba5511} (\bibinfo{year}{2021}).

\bibitem{chen_two-dimensional_2021}
\bibinfo{author}{Chen, Z.} \emph{et~al.}
\newblock \bibinfo{journal}{\bibinfo{title}{Two-{Dimensional} {Superconductivity} at the {LaAlO} 3 / {KTaO} 3 ( 110 ) {Heterointerface}}}.
\newblock {\emph{\JournalTitle{Phys. Rev. Lett.}}} \textbf{\bibinfo{volume}{126}}, \bibinfo{pages}{026802}, \doiprefix\url{10.1103/PhysRevLett.126.026802} (\bibinfo{year}{2021}).

\bibitem{mallik_superfluid_2022}
\bibinfo{author}{Mallik, S.} \emph{et~al.}
\newblock \bibinfo{journal}{\bibinfo{title}{Superfluid stiffness of a {KTaO3}-based two-dimensional electron gas}}.
\newblock {\emph{\JournalTitle{Nature Commun.}}} \textbf{\bibinfo{volume}{13}}, \bibinfo{pages}{4625}, \doiprefix\url{10.1038/s41467-022-32242-y} (\bibinfo{year}{2022}).
\newblock \bibinfo{note}{Number: 1 Publisher: Nature Publishing Group}.

\bibitem{liu_tunable_2023}
\bibinfo{author}{Liu, C.} \emph{et~al.}
\newblock \bibinfo{journal}{\bibinfo{title}{Tunable superconductivity and its origin at {KTaO3} interfaces}}.
\newblock {\emph{\JournalTitle{Nat Commun}}} \textbf{\bibinfo{volume}{14}}, \bibinfo{pages}{951}, \doiprefix\url{10.1038/s41467-023-36309-2} (\bibinfo{year}{2023}).

\bibitem{arnault_anisotropic_2023}
\bibinfo{author}{Arnault, E.~G.} \emph{et~al.}
\newblock \bibinfo{journal}{\bibinfo{title}{Anisotropic superconductivity at {KTaO} $_{\textrm{3}}$ (111) interfaces}}.
\newblock {\emph{\JournalTitle{Sci. Adv.}}} \textbf{\bibinfo{volume}{9}}, \bibinfo{pages}{eadf1414}, \doiprefix\url{10.1126/sciadv.adf1414} (\bibinfo{year}{2023}).

\bibitem{chen_orientation-dependent_2024}
\bibinfo{author}{Chen, X.} \emph{et~al.}
\newblock \bibinfo{journal}{\bibinfo{title}{Orientation-dependent electronic structure in interfacial superconductors {LaAlO3}/{KTaO3}}}.
\newblock {\emph{\JournalTitle{Nat Commun}}} \textbf{\bibinfo{volume}{15}}, \bibinfo{pages}{7704}, \doiprefix\url{10.1038/s41467-024-51969-4} (\bibinfo{year}{2024}).

\bibitem{bareille_two-dimensional_2015}
\bibinfo{author}{Bareille, C.} \emph{et~al.}
\newblock \bibinfo{journal}{\bibinfo{title}{Two-dimensional electron gas with six-fold symmetry at the (111) surface of {KTaO3}}}.
\newblock {\emph{\JournalTitle{Sci Rep}}} \textbf{\bibinfo{volume}{4}}, \bibinfo{pages}{3586}, \doiprefix\url{10.1038/srep03586} (\bibinfo{year}{2015}).

\bibitem{bruno_band_2019}
\bibinfo{author}{Bruno, F.~Y.} \emph{et~al.}
\newblock \bibinfo{journal}{\bibinfo{title}{Band {Structure} and {Spin}–{Orbital} {Texture} of the (111)‐{KTaO} $_{\textrm{3}}$ {2D} {Electron} {Gas}}}.
\newblock {\emph{\JournalTitle{Adv. Electron. Mater.}}} \textbf{\bibinfo{volume}{5}}, \bibinfo{pages}{1800860}, \doiprefix\url{10.1002/aelm.201800860} (\bibinfo{year}{2019}).

\bibitem{mallik_electronic_2023}
\bibinfo{author}{Mallik, S.} \emph{et~al.}
\newblock \bibinfo{journal}{\bibinfo{title}{Electronic band structure of superconducting {KTaO3} (111) interfaces}}.
\newblock {\emph{\JournalTitle{APL Materials}}} \textbf{\bibinfo{volume}{11}}, \bibinfo{pages}{121108}, \doiprefix\url{10.1063/5.0169750} (\bibinfo{year}{2023}).

\bibitem{zhai_large_2023}
\bibinfo{author}{Zhai, J.} \emph{et~al.}
\newblock \bibinfo{journal}{\bibinfo{title}{Large {Nonlinear} {Transverse} {Conductivity} and {Berry} {Curvature} in {KTaO} $_{\textrm{3}}$ {Based} {Two}-{Dimensional} {Electron} {Gas}}}.
\newblock {\emph{\JournalTitle{Nano Lett.}}} \bibinfo{pages}{acs.nanolett.3c03948}, \doiprefix\url{10.1021/acs.nanolett.3c03948} (\bibinfo{year}{2023}).

\bibitem{xu_giant_2024}
\bibinfo{author}{Xu, H.} \emph{et~al.}
\newblock \bibinfo{journal}{\bibinfo{title}{Giant {Tunability} of {Rashba} {Splitting} at {Cation}-{Exchanged} {Polar} {Oxide} {Interfaces} by {Selective} {Orbital} {Hybridization}}}.
\newblock {\emph{\JournalTitle{Advanced Materials}}} \textbf{\bibinfo{volume}{36}}, \bibinfo{pages}{2313297}, \doiprefix\url{10.1002/adma.202313297} (\bibinfo{year}{2024}).
\newblock \bibinfo{note}{\_eprint: https://onlinelibrary.wiley.com/doi/pdf/10.1002/adma.202313297}.

\bibitem{al-tawhid_spin--charge_2025}
\bibinfo{author}{Al-Tawhid, A.~H.} \emph{et~al.}
\newblock \bibinfo{journal}{\bibinfo{title}{Spin-to-charge conversion at {KTaO3}(111) interfaces}}.
\newblock {\emph{\JournalTitle{Applied Physics Letters}}} \textbf{\bibinfo{volume}{126}}, \bibinfo{pages}{091601}, \doiprefix\url{10.1063/5.0247001} (\bibinfo{year}{2025}).

\bibitem{annadi_anisotropic_2013}
\bibinfo{author}{Annadi, A.} \emph{et~al.}
\newblock \bibinfo{journal}{\bibinfo{title}{Anisotropic two-dimensional electron gas at the {LaAlO3}/{SrTiO3} (110) interface}}.
\newblock {\emph{\JournalTitle{Nat Commun}}} \textbf{\bibinfo{volume}{4}}, \bibinfo{pages}{1838}, \doiprefix\url{10.1038/ncomms2804} (\bibinfo{year}{2013}).

\bibitem{pojani_polarity_1999}
\bibinfo{author}{Pojani, A.}, \bibinfo{author}{Finocchi, F.} \& \bibinfo{author}{Noguera, C.}
\newblock \bibinfo{journal}{\bibinfo{title}{Polarity on the {SrTiO3} (111) and (110) surfaces}}.
\newblock {\emph{\JournalTitle{Surface Science}}} \textbf{\bibinfo{volume}{442}}, \bibinfo{pages}{179--198}, \doiprefix\url{10.1016/S0039-6028(99)00911-5} (\bibinfo{year}{1999}).

\bibitem{hua_superconducting_2024}
\bibinfo{author}{Hua, X.} \emph{et~al.}
\newblock \bibinfo{journal}{\bibinfo{title}{Superconducting stripes induced by ferromagnetic proximity in an oxide heterostructure}}.
\newblock {\emph{\JournalTitle{Nat. Phys.}}} \bibinfo{pages}{1--7}, \doiprefix\url{10.1038/s41567-024-02443-x} (\bibinfo{year}{2024}).
\newblock \bibinfo{note}{Publisher: Nature Publishing Group}.

\bibitem{martinez_anisotropic_2023}
\bibinfo{author}{Martínez, E.~A.}, \bibinfo{author}{Dai, J.}, \bibinfo{author}{Tallarida, M.}, \bibinfo{author}{Nemes, N.~M.} \& \bibinfo{author}{Bruno, F.~Y.}
\newblock \bibinfo{journal}{\bibinfo{title}{Anisotropic {Electronic} {Structure} of the {2D} {Electron} {Gas} at the {AlO}/{KTaO3}(110) {Interface}}}.
\newblock {\emph{\JournalTitle{Advanced Electronic Materials}}} \textbf{\bibinfo{volume}{9}}, \bibinfo{pages}{2300267}, \doiprefix\url{10.1002/aelm.202300267} (\bibinfo{year}{2023}).
\newblock \bibinfo{note}{\_eprint: https://onlinelibrary.wiley.com/doi/pdf/10.1002/aelm.202300267}.

\bibitem{johansson_spin_2021}
\bibinfo{author}{Johansson, A.}, \bibinfo{author}{Göbel, B.}, \bibinfo{author}{Henk, J.}, \bibinfo{author}{Bibes, M.} \& \bibinfo{author}{Mertig, I.}
\newblock \bibinfo{journal}{\bibinfo{title}{Spin and orbital {Edelstein} effects in a two-dimensional electron gas: {Theory} and application to {SrTiO} 3 interfaces}}.
\newblock {\emph{\JournalTitle{Phys. Rev. Research}}} \textbf{\bibinfo{volume}{3}}, \bibinfo{pages}{013275}, \doiprefix\url{10.1103/PhysRevResearch.3.013275} (\bibinfo{year}{2021}).

\bibitem{guillet_observation_2020}
\bibinfo{author}{Guillet, T.} \emph{et~al.}
\newblock \bibinfo{journal}{\bibinfo{title}{Observation of large unidirectional rashba magnetoresistance in ge(111)}}.
\newblock {\emph{\JournalTitle{Phys. Rev. Lett.}}} \textbf{\bibinfo{volume}{124}}, \bibinfo{pages}{027201}, \doiprefix\url{10.1103/PhysRevLett.124.027201} (\bibinfo{year}{2020}).

\bibitem{he_bilinear_2018}
\bibinfo{author}{He, P.} \emph{et~al.}
\newblock \bibinfo{journal}{\bibinfo{title}{Bilinear magnetoelectric resistance as a probe of three-dimensional spin texture in topological surface states}}.
\newblock {\emph{\JournalTitle{Nature Phys}}} \textbf{\bibinfo{volume}{14}}, \bibinfo{pages}{495--499}, \doiprefix\url{10.1038/s41567-017-0039-y} (\bibinfo{year}{2018}).

\bibitem{witt_patterning_2023}
\bibinfo{author}{Witt, H.} \emph{et~al.}
\newblock \bibinfo{journal}{\bibinfo{title}{Patterning of superconducting two‐dimensional electron gases based on {AlO} $_{\textrm{ \textit{x} }}$ /{KTaO} $_{\textrm{3}}$ (111) interfaces}}.
\newblock {\emph{\JournalTitle{Advanced Physics Research}}} \textbf{\bibinfo{volume}{2}}, \bibinfo{pages}{2200077}, \doiprefix\url{10.1002/apxr.202200077} (\bibinfo{year}{2023}).

\bibitem{Nord2017}
\bibinfo{author}{Nord, M.}, \bibinfo{author}{Vullum, P.~E.}, \bibinfo{author}{MacLaren, I.}, \bibinfo{author}{Tybell, T.} \& \bibinfo{author}{Holmestad, R.}
\newblock \bibinfo{journal}{\bibinfo{title}{Atomap: a new software tool for the automated analysis of atomic resolution images using two-dimensional gaussian fitting}}.
\newblock {\emph{\JournalTitle{Advanced Structural and Chemical Imaging}}} \textbf{\bibinfo{volume}{3}}, \doiprefix\url{10.1186/s40679-017-0042-5} (\bibinfo{year}{2017}).

\bibitem{ozaki2003variationally}
\bibinfo{author}{Ozaki, T.}
\newblock \bibinfo{journal}{\bibinfo{title}{Variationally optimized atomic orbitals for large-scale electronic structures}}.
\newblock {\emph{\JournalTitle{Physical Review B}}} \textbf{\bibinfo{volume}{67}}, \bibinfo{pages}{155108} (\bibinfo{year}{2003}).

\bibitem{perdew1996generalized}
\bibinfo{author}{Perdew, J.~P.}, \bibinfo{author}{Burke, K.} \& \bibinfo{author}{Ernzerhof, M.}
\newblock \bibinfo{journal}{\bibinfo{title}{Generalized gradient approximation made simple}}.
\newblock {\emph{\JournalTitle{Physical review letters}}} \textbf{\bibinfo{volume}{77}}, \bibinfo{pages}{3865} (\bibinfo{year}{1996}).

\end{thebibliography}

\end{document}